# Identification of Heterogeneous Elastic Properties in Stenosed Arteries: a Numerical Plane Strain Study

ALEXANDRE FRANQUET[(1)], STÉPHANE AVRIL[(1)]*, RODOLPHE LE RICHE[(2)], PIERRE BADEL[(1)]

[(1)] Ecole Nationale Supérieure des Mines de Saint Etienne ; Centre CIS ; Département D2BM-Biomatériaux et Biomécanique ; IFR IFRESIS 143 ; LCG -UMR CNRS 5146 ; 158 Cours Fauriel, 42023 Saint-Étienne Cedex 2, France

[(2)] Ecole Nationale Supérieure des Mines de Saint Etienne ; Institut Fayol, Département DEMO-Décision en Entreprise : Modélisation, Optimisation ; 158 Cours Fauriel, 42023 Saint-Étienne Cedex 2, France

**Abstract**

Assessing the vulnerability of atherosclerotic plaques requires an accurate knowledge of the mechanical properties of the plaque constituents. It is possible to measure displacements *in vivo* inside a plaque using magnetic resonance imaging. An important issue is to solve the inverse problem that consists in estimating the elastic properties inside the plaque from measured displacements. This study focuses on the identifiability of elastic parameters *e.g.* on the compromise between identification time and identification accuracy. An idealised plane strain Finite Element (FE) model is used. The effects of the FE mesh, of the a priori assumptions about the constituents, of the measurement resolution and of the data noise are numerically investigated.

## I. Introduction

Atherosclerosis is a disease where arteries are progressively obstructed, making blood circulation difficult. An atherosclerotic plaque composed essentially of lipids often forms into the artery wall at the arterial tree's bifurcations. It induces an inflammatory response and a local stiffening of the artery (Ross 1999). Ripping-off of atherosclerotic plaques in the carotid arteries is a major cause of mortality in OECD countries. The degree of luminal stenosis, corresponding to the artery's diameter reduction, is commonly used by physicians to decide an endarterectomy (a surgical intervention consisting in removing the plaque). NASCET and ECST studies have shown that patients with large stenosis (70% and more) benefit from carotid endarterectomy (NASCET steering comittee 1991; ECST collaborative group 1991). But it is more difficult to draw conclusions for patients with moderate stenosis. Moreover the phenomenon of artery remodelling makes this geometrical criterion alone insufficient (Casscells *et al.* 2003).

Studies have correlated the arterial compliance with the stroke risk (Kingwell *et al.* 2002). Cheng *et al.* (1993); Ohayon *et al.* (2001); Li *et al.* (2006); Trivedi *et al.* (2007) showed that the vulnerability of an atherosclerotic plaque can be assessed with a maximum stress criterion, which requires an accurate knowledge of the mechanical properties of the different plaque constituents. The estimation of these mechanical properties *in vivo* and the calculation of a stress criterion may be very valuable in helping physicians to evaluate the risk of imminent atherosclerotic plaque's ruptures.

* Corresponding author: avril@emse.fr

Most of the time, the reconstruction of heterogeneous mechanical properties in arteries utilises Finite Element (FE) model updating (Chandran *et al.* 2003; Hamilton *et al.* 2005; Luo *et al.* 2006; Baldewsing *et al.* 2008; Le Floc'h *et al.* 2009).

An inverse problem is solved that consists in finding the parameters of a numerical model so that the predicted deformation fits experimental data (displacements or strains). It is currently not feasible to quantify true strain *in vivo*: arteries are stressed at their free state (Fung 1993; Rachev and Greenwald 2003; Ohayon *et al.* 2007) and their unstressed configuration cannot be assessed *in vivo*. However it is possible to obtain cyclic strains or displacements which are the strains or displacements measured between diastolic and systolic pressures. Ultrasound is one of the techniques that can be employed. De Korte *et al.* (1998) measured radial cyclic strains of a healthy artery thanks to Intravascular Ultrasonography (IVUS). Magnetic Resonance Imaging (MRI) is also a promising method. Draney *et al.* (2002) used a Magnetic Resonance phase contrast (PC-MRI) sequence to obtain radial cyclic strains *in vivo* of healthy arteries.

The *in vivo* identification of heterogeneous mechanical properties of diseased arteries has not been widely investigated in the literature. Baldewsing *et al.* (2008) worked with a 2D parametric finite element model. Elastic properties and contours of the different constituents are recovered one after another from IVUS data. Similarly Le Floc'h *et al.* (2009) created an integrated algorithm where both elastic properties and contours are identified at the same time from IVUS strain data. Computation time can then reach 24 hours for complex geometries.

Inasmuch as a lot of FE models are computed, the first approach to improve the identification time is to work on the FE models. The literature is very scarce on the propagation of errors through the FE mesh choice, besides the study of Le Floc'h (2009). Another approach is to reduce the number of unknown in the identification procedure. It is now well known that arteries are quasi-incompressible (Fung 1993; Black and Hastings 1998) and have a non linear behaviour (Fung 1993; Holzapfel and Weizsacker 1998; Holzapfel 2002; Pena *et al.* 2010). Baldewsing *et al.* (2008) and Le Floc'h *et al.* (2009) considered the mechanical behaviour to be linear elastic and supposed the Poisson's ratios to be known. Moreover the heterogeneities contours of atherosclerotic plaques, which are difficult to discern when data come from IVUS, can be distinguished with MRI (Auer *et al.* 2006; Li *et al.* 2006; Hermus *et al.* 2010).

The originality of this study is to numerically investigate the identifiability of the heterogeneities elastic parameters from simulated PC-MRI data (Draney *et al.* 2002). In particular, the effects of the FE mesh, of the a priori parameters values, of the data spatial resolution and of white noise corrupting experimental data are studied. The analysis focuses on the duration and the accuracy of the identification process.

## II. Materials and methods

### II.1. Finite Element Model

A cross section of an idealised atherosclerotic carotid artery has been modelled in 2D in a manner inspired by Li *et al.* (2008). Dimensions correspond to a 66% stenosis (see Figure 1). One node has been blocked in the $\vec{x}$ and $\vec{y}$ directions, and another has been blocked in the $\vec{y}$ direction to remove rigid body motion. A pressure of 5 kPa is applied uniformly onto the arterial wall, simulating the differential arterial pressure between diastole and systole (see Figure 2). Quasi-static conditions are assumed as the heart rate is approximately 1 Hz. The mechanical behaviour is supposed to be linear elastic, under the hypothesis of plane strain. Three different materials are defined:

(1) Healthy tissue.

(2) A lipidic core composed essentially of fat.

(3) Diseased tissue including the fibrous cap and a part of the infected media, stiffer than the healthy tissue.

Hypotheses and assumptions of the numerical model are discussed in Section IV.

## II.2. Reference data

Reference data serve as pseudo-experimental data. The behaviour of arteries is supposed to be quasi-incompressible. The elastic properties of the three materials are set to the following values:

(1) Healthy tissue: $E_{HT} = 600kPa$ ; $\nu_{HT} = 0.49$

(2) Lipidic core: $E_{LC} = 10kPa$ ; $\nu_{LC} = 0.49$

(3) Diseased tissue: $E_{DT} = 800kPa$ ; $\nu_{DT} = 0.49$

The reference data are the result of the computation of a FE model with about 130 000 CPE8H elements (hybrid plane strain quadratic quadrangular elements). This element allows to treat the hydrostatic pressure as an independent variable and is recommended for quasi-incompressible cases (ABAQUS 6.8 2008). The displacement field solution is then linearly interpolated on a regular grid of step S, which simulates a PC-MRI output with the voxel size S (see Figure 3) (Draney *et al.* 2002).

## II.3. Inverse approach

An inverse approach consists in finding the parameters of a system, knowing its response. The principle is to minimise a distance $J_2$ (see Equation 1) with an optimisation algorithm.

$$\min_{\vec{\theta}} . J_2(\vec{\theta}) \tag{1}$$

where

$$J_2(\vec{\theta}) = \frac{1}{2}\sum_{i=0}^{N}(U^i_{reference} - u^i(\vec{\theta}))^2 \tag{2}$$

$N$ : Number of interpolation nodes (grid nodes).

$(\vec{\theta})$ : Vector of parameters, Young's moduli and Poisson's ratios depending on the test case considered (see Section II.4).

$u^i(\vec{\theta})$ : Displacement from a FE simulation interpolated at the grid node *i*.

$U^i_{reference}$ : Reference displacement at the grid node *i*.

A Levenberg-Marquardt algorithm with bounds handling (Marquardt 1963; Guyon and Le Riche 2000) is used to recover the elastic properties (see Figure 4). This algorithm needs gradients of $U^i_{x,y}$ with respect to $\vec{\theta}$ which are estimated by finite differences. Two termination criteria are set up:

$J_2(\vec{\theta}) \leq \varepsilon_a = 10^{-7}$ : Accuracy on $J_2$ reached.

$\|\Delta\vec{\theta}\| \leq \varepsilon_b = 10^{-25}$ : step $\|\Delta\vec{\theta}\|$ too small. No more improvement is expected.

The geometry of the FE model used is similar to the one used for generating reference displacement (see Figure 1). The displacements are always projected on the same grid as the reference data (see Figure 3(b)) and using the same procedure (see Section II.2).

## II.4. Definition of numerical experiments

This section defines four series of tests which aim at studying the identifiability of mechanical properties. Starting from a default configuration (see Table 1), different parameters of the FE model and of the identification procedure are changed as explained hereafter. It is possible to measure the quality of an identification simply by comparing the identified mechanical properties (see Figure 4) to the target values defined in Section II.2. The total computation time is also observed. The influence of the FE mesh, the *a priori* Poisson's ratios, the grid' step size and the data noise are studied as explained hereafter.

### II.4.1. Effect of the FE mesh

The first step is to choose appropriate meshes for subsequent tests. Two elements are tested: the CPE6 (plane strain quadratic triangular element, used in Le Floc'h (2009)) and the CPE8H. Four meshes made of 1000, 5000, 15000 and 40000 elements are investigated.

### II.4.2. Effect of the a priori Poisson's ratios

As explained in Section I the Poisson's ratios are supposed to be known. However the choice of Poisson's ratios can have an influence on the Young's moduli identification. To study this sensitivity, Poisson's ratios are voluntarily unsettled during the identification procedure:

$$\{v_{HT}, v_{LC}, v_{DT},\} = \{\{0.45, 0.45, 0.45\}, \{0.48, 0.48, 0.48\}, \{0.49, 0.49, 0.49\} \{0.499, 0.499, 0.499\}\} \quad (3)$$

### II.4.3. Effect of the grid step size

The grid step size is directly linked to the MRI spacial resolution. Current limitations of MRI devices are prone to affect the identification quality. The influence of MRI accuracy on Young's moduli recovery is investigated by interpolating the reference data on a regular grid of step sizes S taken as:

$$S = \{1\ mm, 0.5\ mm, 0.25\ mm, 0.125\ mm\} \quad (4)$$

### II.4.4. Effect of a white Gaussian noise

The effect of a white Gaussian noise added to the reference data is studied because experimental data always contain noise. The reference displacements are defined as:

$$\overrightarrow{U^x_{reference}} = \overrightarrow{U^x_{reference}} + \sigma \times \overrightarrow{R_1} \quad (5)$$

$$\overrightarrow{U^y_{reference}} = \overrightarrow{U^y_{reference}} + \sigma \times \overrightarrow{R_2} \quad (6)$$

With:

$\sigma = 3\% \times \left\|\overrightarrow{U_{reference}}\right\|$ : standard deviation

$\overrightarrow{R_1}$ and $\overrightarrow{R_2}$ : random vectors following a standard normal law $N(0, I)$ where $I$ is the $N_{interpolation\ nodes} \times N_{interpolation\ nodes}$ identity matrix.

A series of twenty different random noise vectors are added to the reference data in order to perform twenty different identifications. The default identification parameters are used except for the noise level (see Table 1). The mean and standard deviation of the errors on the identified values and the identification time are calculated.

## III. Results

### III.1. Effect of the FE mesh

Results are reported in Table 2. The identification with the CPE6 element always yields artery Young's moduli with an error lower than 2.4%. The errors of CPE8H identifications are always lower than 0.7% whatever the number of elements is. Five identifications terminated because sufficient accuracy on $J_2$ was reached ($\varepsilon_a$) and three terminated because of too small a step size ($\varepsilon_b$). It is remarkable that for low number of elements, the $J_2$ index has not reached the limit $\varepsilon_a$ although the identification quality is still better than 2.4%.

This test leads to the selection of one mesh for each element type for the next numerical experiments. The best compromise between time and accuracy for the CPE8H element type is obtained for 5000 elements as explained hereafter. Concerning the CPE6 element type it appears that 15000 elements must be the best compromise. However we noticed that the algorithm reaches an accuracy of 1% after 346 seconds with 5000 elements (see Figure 5). Thus the two following meshes are selected as they are the best compromise between time and accuracy (see Figure 5):

- ❖ CPE6 5000: Although the accuracy on $J_2$ has not been reached, the algorithm finds a vector $\vec{\theta}_{final}$ after only 346 seconds:

$$\vec{\theta}_{final} = \{E_{HA} = 606 \ kPa, \ E_{LC} = 9.99 \ kPa, \ E_{DT} = 799 \ kPa\}$$

- ❖ CPE8H 5000: The identification quality is good (<1% on all Young's moduli), and the identification total time (557 seconds) is the lowest among identifications which converged according to the $J_2$ stopping criterion $\varepsilon_a$. The identified Young's moduli are:

$$\vec{\theta}_{final} = \{E_{HA} = 604 \ kPa, \ E_{LC} = 10.04 \ kPa, \ E_{DT} = 800 \ kPa\}$$

### III.2. Effect of the *a priori* Poisson's ratios

Results are reported in Table 3. Poisson's ratios of 0.45 lead to errors of + 8% on the healthy artery Young's modulus, and a + 80% overestimation on the lipidic core Young's modulus for both meshes, whereas Poisson's ratios of 0.48 lead to errors of respectively + 5% and + 37%. Poisson's ratios of 0.499 significantly reduce $E_{HA}$ (-35%) and $E_{LC}$ (-87%).

### III.3. Effect of the grid step size

The tests involving different grid step sizes are reported in Table 4. The results are very similar in terms of accuracy for the two types of elements. With the larger step size (1 mm), the error on the healthy artery Young's modulus is 9% and the error on the lipidic core is 8%. The identification time decreases when the step size decreases (excepted for CPE6 / 0.125 mm which terminated with a different criterion, $\|\Delta\vec{\theta}\| \leq 10^{-25}$). Note that with a given step size, for instance 1 mm, the CPE6 identification is twice as fast as that with CPE8H elements.

### III.4. Effect of white Gaussian noise

Results are reported in Table 5. Twenty identifications have been performed from the same default initial parameters (see Table 1) for the two meshes. Twenty random noise vectors were added to reference displacements. The added noise affects both element types in the same marginal way in terms of Young's moduli standard deviations (respectively 0.21%, 0.30%, 0.03%).

## IV. Discussion

### IV.1. Stopping criterion

To our knowledge the effects of the criteria ($\varepsilon_a$ and $\varepsilon_b$) on the identification of artery's mechanical properties have never been investigated. Our trials show the relative significance of the $J_2$ value since an identification with a coarser mesh may have a high $J_2$ value leading to the same identification quality than using a finer mesh. The Figure 6 shows that the minimum of $J_2$ is found at the value $E_{lipidic\ core}$ = 10 kPa (which is the reference value for this parameter) irrespectively of the mesh while the minimum value of $J_2$ depends on the chosen mesh.

Here, both selected models have 5000 elements, which can be compared to the 1152 triangular elements of Kallel and Bertrand (1996), the 2000 quadrangular elements of Khalil *et al.* (2006), the 4100 triangular elements of Fehrenbach *et al.* (2006) and the 15000 quadratic triangular elements of Le Floc'h *et al.* (2009). We show that an appropriate choice of FE model can save a substantial amount of computation time. Indeed, a FE model with only 5000 quadratic triangular elements is sufficient to obtain an identification accuracy lower than 1% while the identification time is potentially three times as fast as the identification with a 15000 quadratic quadrangles model (considering a better choice of termination criterion). In our opinion it represents a good compromise between time and accuracy. A less stringent choice of threshold for the second termination criterion $\varepsilon_b$ can significantly reduce the computation time when this criterion is predominant.

### IV.2. Model assumptions

#### IV.2.1. Model geometry

Atherosclerotic plaques are classified in categories (Cai *et al.* 2002). The artery modelled is an idealised type IV-V diseased artery where a large lipid pool is surrounded by fibrotic tissue. It should be kept in mind that MRI can distinguish different constituents, and not directly the different mechanical properties. However here we assume that the MRI data would provide the segmentation as presented in Figure 1.

#### IV.2.2. Surrounding tissues

Carotid arteries are encircled by soft tissue. Liu *et al.* (2007) already worked on the impact of such media on the arterial wall strains and highlighted its crucial role. Le Floc'h (2009) also studied the influence of an elastic medium surrounding the artery on the mechanical properties identification. It does not have an influence in our case because we use reference data and similar FE models between the pseudo-experimentation and the computations for the identification. But surrounding tissues should be considered in practice.

#### IV.2.3. Constitutive law

Even though the behaviour of arteries is known to be non-linear (for example Holzapfel (2002)), Khalil *et al.* (2006) underlined that the incompressible linear elastic stress-strain constitutive relations under isotropic and plane-strain assumptions are sufficient to model the behaviour of soft tissue undergoing small and quasi-static deformations. Moreover, Chabanas *et al.* (2004) pointed out that linear elasticity is acceptable for FE simulations of soft human tissue in the physiological range of pressure. The constitutive law chosen in this study is a linear approximation of the stress and strain variations in the diastole to systole range of loading. The elastic parameters can be interpreted as tangent moduli.

#### IV.2.4. Quasi-incompressibility

In this problem the Poisson's ratios are unknown as are the Young's moduli. Our results (see Section III.2) show that a bad estimation of the Poisson's ratio has a major influence on the identification quality in particular for the lipidic core whereas the Young's moduli of the healthy artery and of the diseased tissue are mainly affected in the case of an overestimation of the Poisson's ratios. In Table 6 we compare the identification times and accuracies of optimisations when Poisson's ratios are set to 0.49 (identification of the three Young's moduli) against the identification times and accuracies when Poisson's ratios are three additional parameters to identify (identification of six parameters). The identification of six parameters provides an accuracy below 0.07% on the Young's moduli and below 0.35% on the Poisson's ratios. This table highlights that it is possible to save about three times the identification time by setting the Poisson's ratios. The accuracy on the Young's moduli is then negligibly impacted (<1%). This means that in a more general case the estimation of the Poisson's ratios is a key point: either the Poisson's ratios are accurately known and can be set in the identification procedure or these ratios must be identified together with the Young's moduli at the expense of an additional computational cost.

Starting from an heterogeneous coronary artery model with similar mechanical hypothesis and two different materials (a diseased tissue and a lipidic core), Le Floc'h (2009) kept a Poisson's ratio set to 0.49 in his identification method, but modified the reference data by changing the diseased tissue Poisson's ratio. Surprisingly it appeared that the method overestimated the Young's modulus of the artery when the Poisson's ratio is overestimated in the numerical model. On the contrary, our study reveals that the algorithm tends to soften the healthy artery and the lipidic core to compensate an overestimation of Poisson's ratios and vice versa. Le Floc'h (2009) worked with a fully integrated method where heterogeneities contours are identified, while the present study keeps the contours fixed and focuses only on the identification issues. This could explain the difference.

### IV.2.5. Pre stress

An initial stress is induced by the blood pressure. The lowest pressure encountered in arteries is approximately 80 mmHg which corresponds to 10.66 kPa. Ohayon *et al.* (2007) considered the stress at the lumen border to be zero at diastole. Residual stresses are present inside the artery at free state as it has been shown in Fung (1993); Rachev and Greenwald (2003); Ohayon *et al.* (2007). It is currently difficult to assess this stress *in vivo*, although Ohayon *et al.* (2005) mentioned a method to estimate the initial configuration of a patient's coronary artery. Not taking into account these pre stresses can have a major influence on the final deformed artery (Alastrue *et al.* 2010), but it does not affect the current work since numerical reference data are considered and only the tangent moduli are estimated.

### IV.3. Origin and quality of data

The grid step size has no visible effect on the accuracy of the identified Young's moduli from the step size of 0.5 mm downwards. However, in an experimental context, data can be noisy and affect the identification procedure. A large amount of data may be the key to reduce the identification time and to improve the accuracy. The results reported in Table 4, although deprived of any noise, hint at this trend. White Gaussian noise added to the reference data does not affect the identification time nor accuracy. Note that it is difficult to define a noise -3% here- which reflects a real experimental noise as we work with displacements as opposed to measures such as pixel intensity.

## V. Conclusion

In this study, we have modelled an idealised plane strain atherosclerotic carotid artery with three different materials and have identified its tangent elastic properties with a Levenberg-Marquardt algorithm. We focused on how the numerical implementation affects identifiability of the Young's moduli where the identifiability describes here the compromise between moduli accuracy and computation time. The effects of the FE mesh refinement, of the Poisson's ratios, of data resolution, and of noisy data have been studied.

A FE model of 5000 quadratic triangular elements was shown to be sufficient for reaching an accuracy of 1% on the identified Young's moduli, although a finer mesh will better reflect the analytical displacements.

This leads to a second result: the classical index $J_2$ used in optimisation strategies depends on the mesh choice and should not be the only identification criterion. A second termination criterion corresponding to the change of mechanical properties between consecutive iterations is necessary to allow, in conjunction with a light FE mesh, a reduced computation time.

Our results also emphasize the importance of a correct estimation of the Poisson's ratios which can significantly alter the results.

This study mainly addressed errors arising from the model and from the inversion method. Another source of difficulties is the experimental data and its simulation which has been questioned by superimposing white Gaussian noise onto the reference displacements. Future work will focus on obtaining more realistic data by simulating a real MRI output and finally by working on real experimental data.

## Acknowledgements

This work is part of the Imandef Project (*in vivo* mechanical identification of tissues using medical imaging, grant ANR-08-JCJC-0071) funded by the ANR (French National Research Agency).

**Figures**

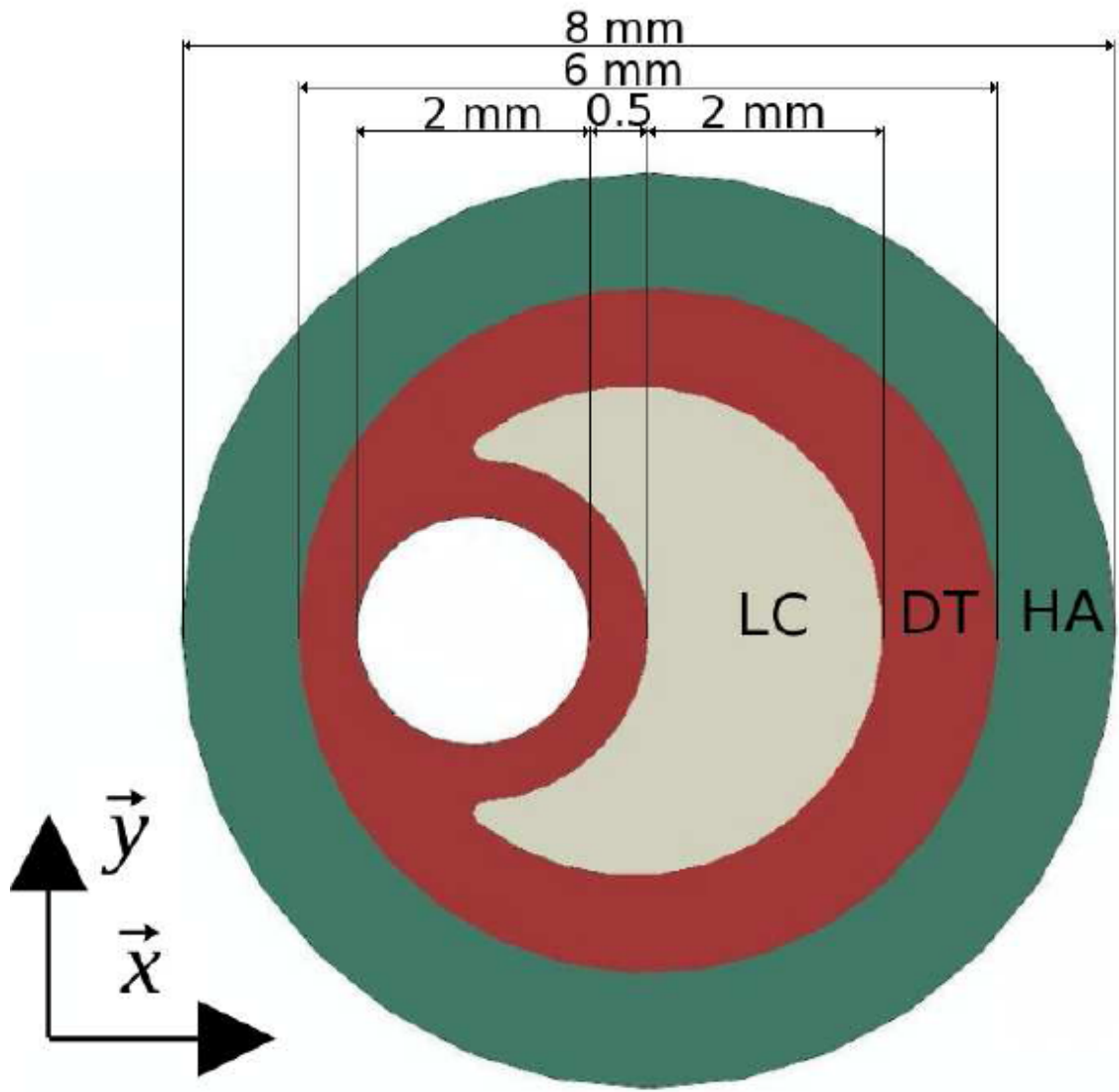

*Figure 1: Dimensions and materials of the artery model. The lumen's initial shape is represented by the healthy artery (HA). Atherosclerosis induces some diseased tissue (DT) which modifies the blood circulation. The lipidic core (LC) is separated from the lumen by a fibrous cap included in the DT.*

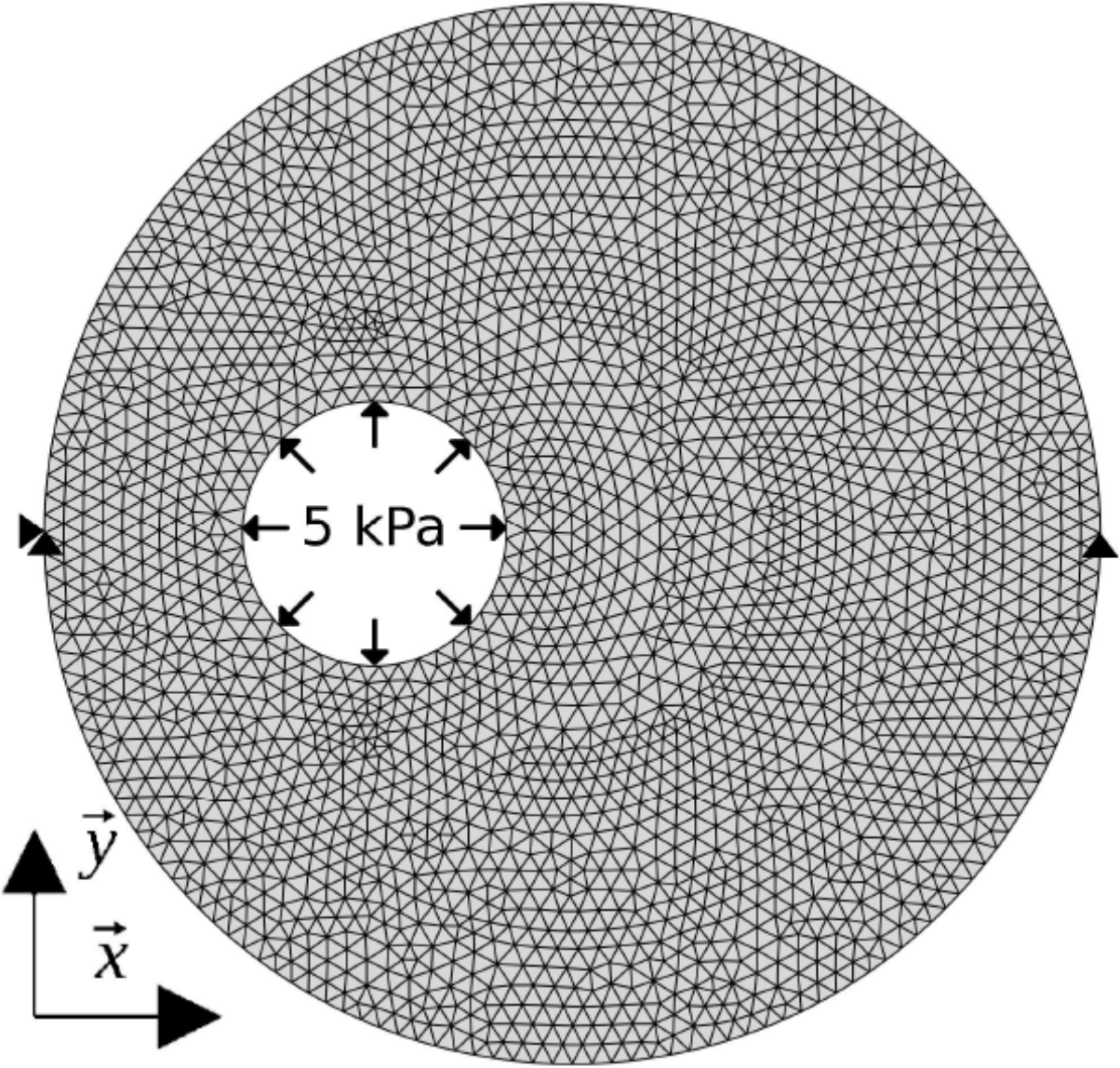

*Figure 2: Boundary conditions of the artery Finite Element model.*

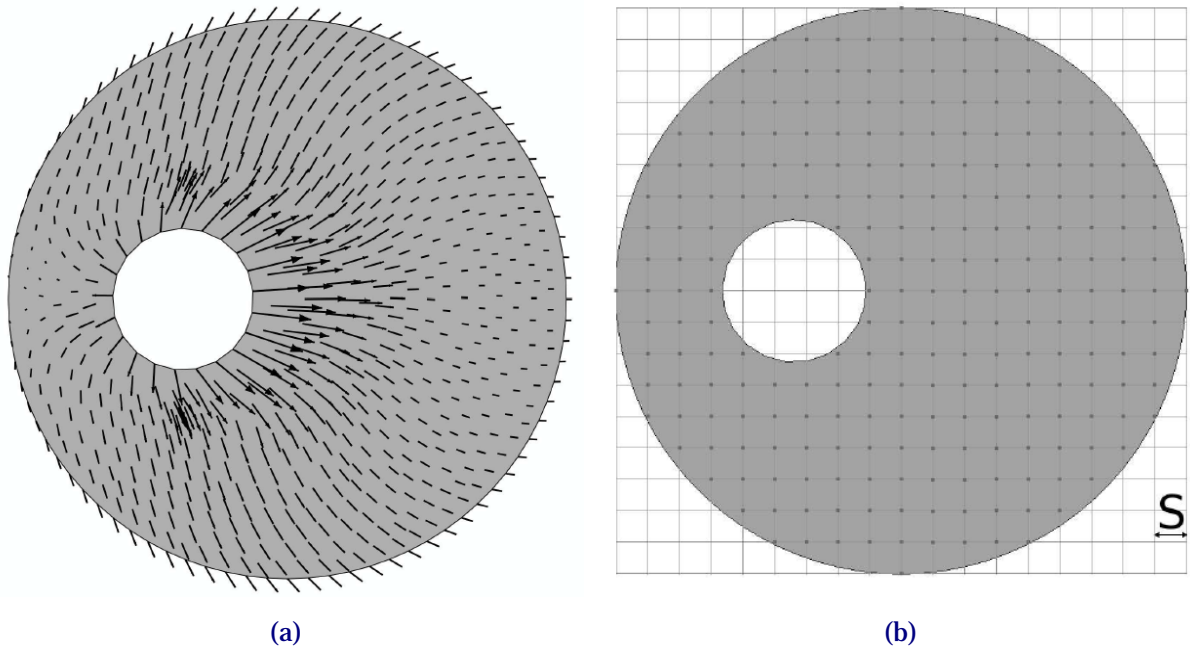

(a)          (b)

*Figure 3: Displacements from the FE computation (a) are linearly interpolated on the grid nodes (b). (a): Displacements at the nodes of the FE computation. (b): Interpolation grid with a voxel size S.*

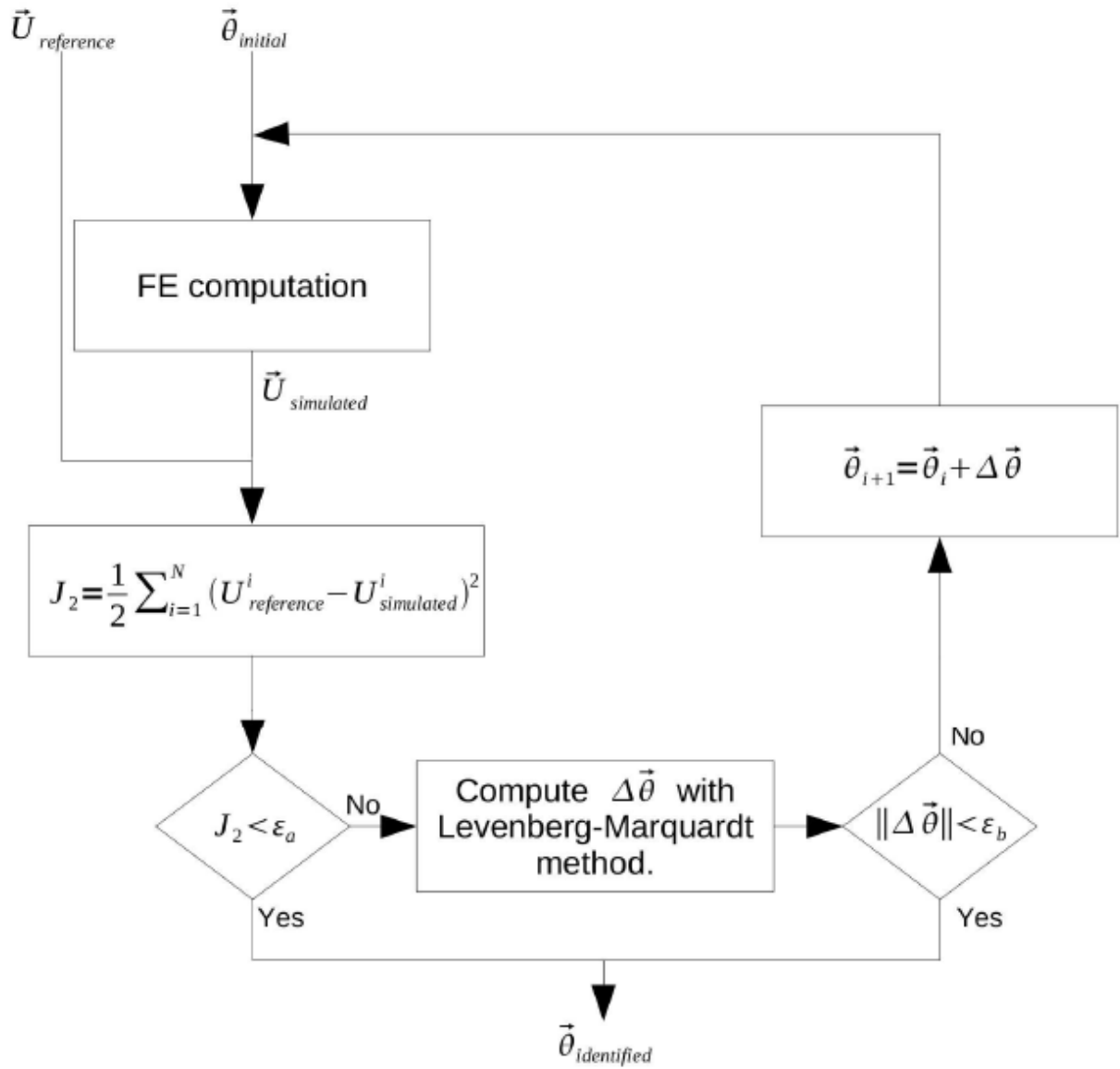

Figure 4: Flow chart of the inverse method.

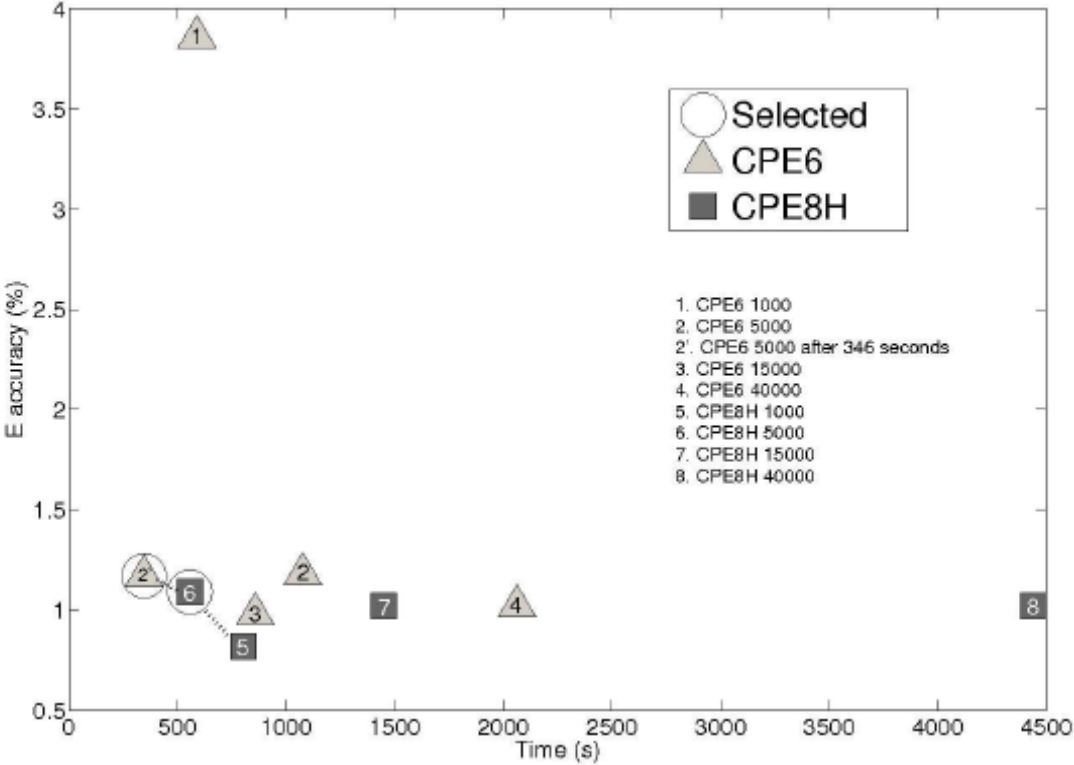

*Figure 5: Pareto frontier accuracy/time for the different meshes. Accuracy is defined as the sum of the absolute values of errors on each Young's modulus.*

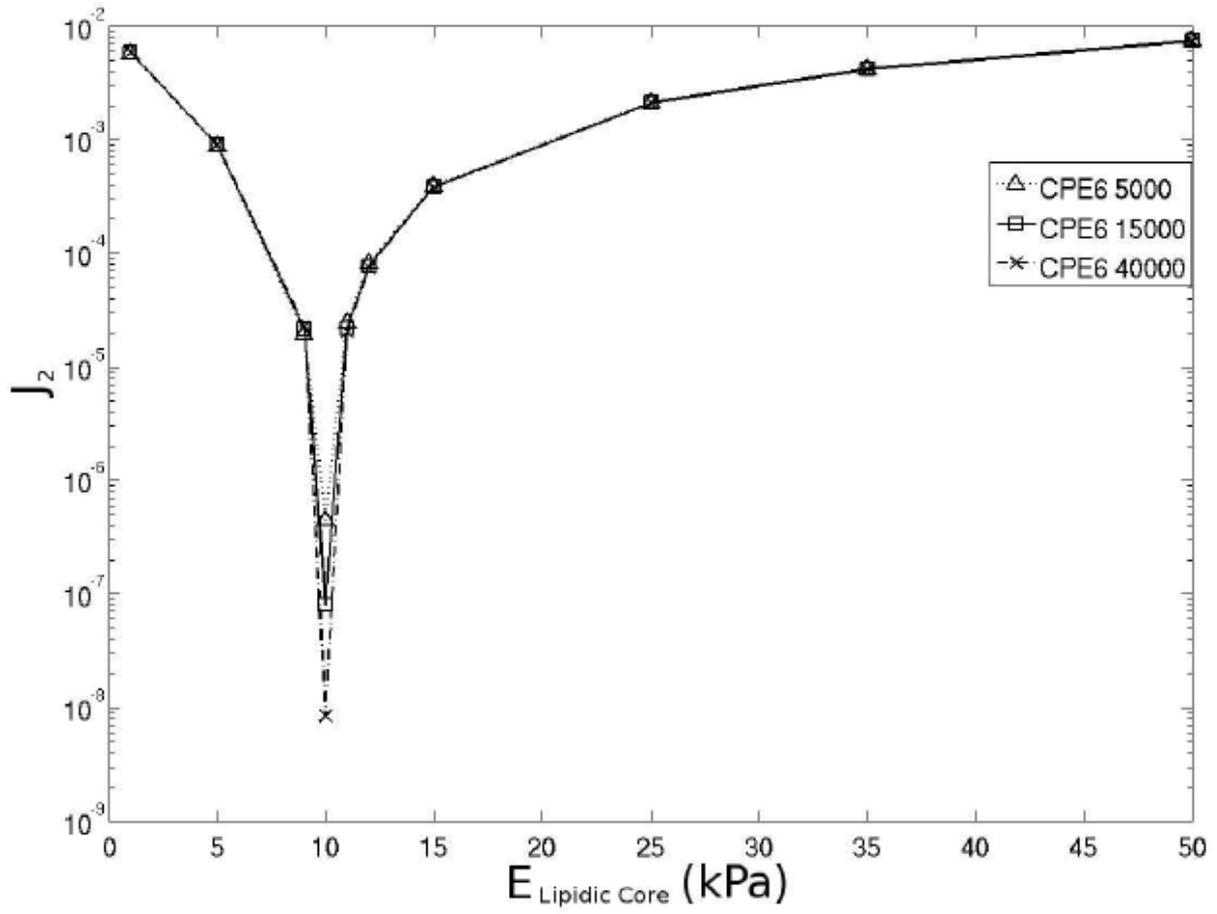

*Figure 6: $J_2$ criterion as a function of $E_{lipidic\ core}$ for a CPE6 mesh with 5000, 15000 or 40000 elements. Others mechanical properties are set to the reference values.*

# Tables

*Table 1: Default parameters*

| Initial vector of parameters | $\bar{\theta} = \{E_{HT} = 1000\ kPa,\ E_{LC} = 100\ kPa,\ E_{DT} = 1200\ kPa\}$ |
|---|---|
| Mesh types | CPE6 5000 and CPE8H 5000 |
| Set of Poisson ratios | $\{v_{HT}, v_{LC}, v_{DT},\} = \{0.49, 0.49, 0.49\}$ |
| Grid step for the synthetic data | 0.125 mm |
| Noise level added to synthetic data | 0% |

*Table 2: Identification time and accuracy of optimisation procedures with respect to the size and the type of the FR mesh*

| Mesh type | | $E_{HA}$ (kPa) | % $E_{HA}$ (kPa) | $E_{LC}$ (kPa) | % $E_{LC}$ (kPa) | $E_{DT}$ (kPa) | % $E_{DT}$ (kPa) | Time (s) 1 FE | Termination Total criterion |
|---|---|---|---|---|---|---|---|---|---|
| References | | 600 | | 10.00 | | 800 | | | |
| CPE6 | 1000 | 607 | 1.04% | 9.77 | -2.33% | 798 | -0.29% | 0.38 | 590 | $\|\Delta\vec{\theta}\| \leq 10^{-25}$ |
| CPE6 | 5000 | 603 | 0.55% | 9.94 | -0.56% | 799 | -0.07% | 1.19 | 1038 | $\|\Delta\vec{\theta}\| \leq 10^{-25}$ |
| CPE6 | 15000 | 604 | 0.68% | 10.02 | 0.24% | 799 | -0.06% | 3.99 | 858 | $J_2(\vec{\theta}) \leq 10^{-7}$ |
| CPE6 | 40000 | 603 | 0.53% | 10.04 | 0.44% | 800 | -0.05% | 10.66 | 2063 | $J_2(\vec{\theta}) \leq 10^{-7}$ |
| CPE8H | 1000 | 602 | 0.43% | 9.97 | -0.33% | 800 | -0.06% | 0.84 | 803 | $\|\Delta\vec{\theta}\| \leq 10^{-25}$ |
| CPE8H | 5000 | 604 | 0.64% | 10.04 | 0.39% | 800 | -0.06% | 3.83 | 557 | $J_2(\vec{\theta}) \leq 10^{-7}$ |
| CPE8H | 15000 | 604 | 0.54% | 10.04 | 0.43% | 800 | -0.05% | 12.39 | 1149 | $J_2(\vec{\theta}) \leq 10^{-7}$ |
| CPE8H | 40000 | 603 | 0.47% | 10.05 | 0.51% | 800 | -0.04% | 44.63 | 4440 | $J_2(\vec{\theta}) \leq 10^{-7}$ |

Table 3: Identification time and accuracy of optimisation procedures when Poisson ratios are set to wrong values in the identification process.

| Mesh type | | $\upsilon_{HA}$ | $\upsilon_{LC}$ | $\upsilon_{DT}$ | $E_{HA}$ (kPa) | % $E_{HA}$ (kPa) | $E_{LC}$ (kPa) | % $E_{LC}$ (kPa) | $E_{DT}$ (kPa) | % $E_{DT}$ (kPa) | Total time (s) | Termination criterion |
|---|---|---|---|---|---|---|---|---|---|---|---|---|
| References | | 0.49 | 0.49 | 0.49 | 600 | | 10.00 | | 800 | | | |
| | | | | | | | | | | | | |
| CPE6 | 1000 | 0.45 | 0.45 | 0.45 | 650 | 8.26% | 17.90 | 78.99% | 804 | 0.45% | 1017 | $\|\Delta\vec{\theta}\| \leq 10^{-25}$ |
| | 5000 | 0.48 | 0.48 | 0.48 | 633 | 5.47% | 13.65 | 36.48% | 793 | -0.87% | 1002 | $\|\Delta\vec{\theta}\| \leq 10^{-25}$ |
| | 15000 | 0.49 | 0.49 | 0.49 | 603 | 0.55% | 9.94 | -0.56% | 799 | -0.07% | 1038 | $\|\Delta\vec{\theta}\| \leq 10^{-25}$ |
| | 40000 | 0.499 | 0.499 | 0.499 | 394 | -34.66% | 1.23 | -87.74% | 844 | 5.53% | 1187 | $\|\Delta\vec{\theta}\| \leq 10^{-25}$ |
| | | | | | | | | | | | | |
| CPE8H | 1000 | 0.45 | 0.45 | 0.45 | 648 | 8.03% | 17.92 | 79.22% | 804 | 0.47% | 1579 | $\|\Delta\vec{\theta}\| \leq 10^{-25}$ |
| | 5000 | 0.48 | 0.48 | 0.48 | 631 | 5.21% | 13.69 | 36.88% | 793 | -0.84% | 1620 | $\|\Delta\vec{\theta}\| \leq 10^{-25}$ |
| | 15000 | 0.49 | 0.49 | 0.49 | 604 | 0.64% | 10.04 | 0.39% | 800 | -0.06% | 557 | $J_2(\vec{\theta}) \leq 10^{-7}$ |
| | 40000 | 0.499 | 0.499 | 0.499 | 387 | -35.43% | 1.22 | -87.82% | 845 | 5.68% | 2063 | $\|\Delta\vec{\theta}\| \leq 10^{-25}$ |

*Table 4: Identification time and accuracy of optimisation procedures when reference data are interpolated on grids with different steps sizes*

| Mesh type | S (mm) | $E_{HA}$ (kPa) | % $E_{HA}$ | $E_{LC}$ (kPa) | % $E_{LC}$ | $E_{DT}$ (kPa) | % $E_{DT}$ (kPa) | Total time (s) | Termination criterion |
|---|---|---|---|---|---|---|---|---|---|
| References |  | 600 |  | 10.00 |  | 800 |  |  |  |
| CPE6 5000 | 1 | 655 | 9.12% | 10.76 | 7.62% | 795 | -0.64% | 462 | $J_2(\vec{\theta}) \leq 10^{-7}$ |
|  | 0.5 | 607 | 1.16% | 10.01 | 0.11% | 799 | -0.12% | 443 | $J_2(\vec{\theta}) \leq 10^{-7}$ |
|  | 0.25 | 606 | 1.04% | 10.00 | -0.01% | 799 | -0.11% | 392 | $J_2(\vec{\theta}) \leq 10^{-7}$ |
|  | 0.125 | 603 | 0.55% | 9.94 | -0.56% | 799 | -0.07% | 1038 | $\|\Delta\vec{\theta}\| \leq 10^{-25}$ |
| CPE8H 5000 | 1 | 654 | 8.99% | 10.83 | 8.26% | 795 | -0.63% | 802 | $J_2(\vec{\theta}) \leq 10^{-7}$ |
|  | 0.5 | 605 | 0.87% | 10.83 | 0.65% | 799 | -0.07% | 753 | $J_2(\vec{\theta}) \leq 10^{-7}$ |
|  | 0.25 | 604 | 0.74% | 10.06 | 0.50% | 799 | -0.07% | 653 | $J_2(\vec{\theta}) \leq 10^{-7}$ |
|  | 0.125 | 604 | 0.64% | 10.05 | 0.39% | 800 | -0.06% | 557 | $J_2(\vec{\theta}) \leq 10^{-7}$ |

Table 5: Means and standard deviations of time and accuracy of twenty identifications for each mesh when 3% random Gaussian noise is added into the reference data.

| Mesh type | | $E_{HA}$ (kPa) | % $E_{HA}$ (kPa) | $E_{LC}$ (kPa) | % $E_{LC}$ (kPa) | $E_{DT}$ (kPa) | % $E_{DT}$ (kPa) | Total time (s) | Termination criterion |
|---|---|---|---|---|---|---|---|---|---|
| References | | 600 | | 10.00 | | 800 | | | |
| CPE6 5000 | Mean | 603 | 0.52% | 9.93 | -0.65% | 799 | -0.08% | 1114 | $\|\Delta\vec{\theta}\| \leq 10^{-25}$ |
| | Std. dev. | 1.28 | 0.21% | 0.03 | 0.26% | 0.25 | 0.03% | 48 | |
| CPE8H 5000 | Mean | 601 | 0.17% | 9.98 | -0.20% | 800 | -0.03% | 1768 | $\|\Delta\vec{\theta}\| \leq 10^{-25}$ |
| | Std. dev. | 1.28 | 0.21% | 0.03 | 0.26% | 0.25 | 0.03% | 61 | |

*Table 6: Comparison between the identification time and accuracy of the optimisation procedures when the Poisson's ratios are set to 0.49 (3 parameters identification) or left as independent variables (6 parameters)*

| Mesh type | Number of parameters | $E_{HA}$ (kPa) | % $E_{HA}$ (kPa) | $\upsilon_{HA}$ | % $\upsilon_{HA}$ | $E_{LC}$ (kPa) | % $E_{LC}$ (kPa) | $\upsilon_{LC}$ | % $\upsilon_{LC}$ | $E_{DT}$ (kPa) | % $E_{DT}$ (kPa) | $\upsilon_{DT}$ | % $\upsilon_{DT}$ | Total time (s) | Termination criterion |
|---|---|---|---|---|---|---|---|---|---|---|---|---|---|---|---|
| References | | 600 | | 0.49 | | 10.00 | | 0.49 | | 800 | | 0.49 | | | |
| CPE6 | 3 | 603 | 0.55% | 0.49 | set | 9.94 | -0.56% | 0.49 | set | 799 | -0.07% | 0.49 | set | 1038 | $\|\Delta\vec{\theta}\| \leq 10^{-25}$ |
| 5000 | 3 (after 346s) | 606 | 0.95% | 0.49 | set | 9.99 | -0.10% | 0.49 | set | 799 | -0.06% | 0.49 | set | 346 | |
| | 6 | 600 | 0.07% | 0.49 | 0.35% | 9.95 | -0.52% | 0.49 | 0.01% | 800 | 0.03% | 0.49 | -0.29% | 1088 | $J_2(\vec{\theta}) \leq 10^{-7}$ |
| CPE8H | 3 | 604 | 0.64% | 0.49 | set | 10.04 | 0.39% | 0.49 | set | 800 | -0.06% | 0.49 | set | 557 | $J_2(\vec{\theta}) \leq 10^{-7}$ |
| 5000 | 6 | 600 | 0.03% | 0.49 | 0.26% | 10.00 | 0.01% | 0.49 | 0.00% | 800 | 0.05% | 0.49 | 0.22% | 1404 | $J_2(\vec{\theta}) \leq 10^{-7}$ |